\documentclass[aps,pra,showpacs,twocolumn,amsmath,amssymb]{revtex4}
\usepackage{bm}
\usepackage{graphicx}

\newcommand{\uu}[1]{\underline{#1}}
\newcommand{\pp}[1]{\phantom{#1}}
\newcommand{\be}{\begin{eqnarray}}
\newcommand{\ee}{\end{eqnarray}}

\newcommand{\ba}{\begin{array}}
\newcommand{\ea}{\end{array}}

\newcommand{\sinc}{{\,\rm sinc\,}}

\begin{document}

\title{
Degree of entanglement as a physically ill-posed problem: The case of entanglement with vacuum}
\author{Marcin Paw{\l}owski and Marek Czachor}

\affiliation{
Katedra Fizyki Teoretycznej i Metod Matematycznych\\
Politechnika Gda\'nska, 80-952 Gda\'nsk, Poland}

\begin{abstract}
We analyze an example of a photon in superposition of different modes, and ask what is the degree of their entanglement with vacuum. 
The problem turns out to be ill-posed since we do not know which representation of the algebra of canonical commutation relations (CCR) to choose for field quantization. Once we make a choice, we can solve the question of entanglement unambiguously. So the difficulty is not with mathematics, but with physics of the problem. In order to make the discussion explicit we analyze from this perspective a popular argument based on a photon leaving a beam splitter and interacting with two two-level atoms. We first solve the problem algebraically in Heisenberg picture, without any assumption about the form of representation of CCR. Then we take the $\infty$-representation and show in two ways that in two-mode states the modes are maximally entangled with vacuum, but single-mode states are not entangled. Next we repeat the analysis in terms of the representation of CCR taken from Berezin's book and show that two-mode states do not involve the mode-vacuum entanglement. Finally, we switch to a family of reducible representations of CCR recently investigated in the context of field quantization, and show that the entanglement with vacuum is present even for single-mode states. Still, the degree of entanglement is here difficult to estimate, mainly because there are $N+2$ subsystems, with $N$ unspecified and large.
\end{abstract}


\maketitle

\section{Introduction}

One of the fundamental postulates of quantum theory is that symmetry groups are represented unitarily. In principle, all unitary representations of groups are allowed. The choice of a concrete represenatation is dictated by the values of `quantum numbers'. Irreducibility of representations is not fundamental. Multiparticle systems are usually described by reducible representations since a tensor product of two irreducible representations is typically reducible. Restriction of physical analysis to an irreducible `block' is not always justified, and superpositions of states belonging to different blocks may be acceptable. 

When it comes to quantum fields the main symmetry of interest is the one associated with the harmonic-oscillator Lie algebra of canonical commutation relations (CCR). The field is often assumed to consist of many, perhaps even infinitely many, oscillators. One often hears that the field is a system with infinitely many degrees of freedom. Another way of phrasing this is that the CCR corresponds to a system with  infinitely many degrees of freedom, a fact implying by von Neumann's theorem that it has infinitely many inequivalent irreducible representations. So even if we impose irreducibility as a constraint, which representation should we choose? Do all of them imply the same physics? Maybe different experimental configurations require different representations? Moreover, if the field is indeed a multi-oscillator system, why should we impose the irreducibility constraint? -- and so on.

For the moment there is no clear answer to any of these questions, although an experiment that may distinguish at least between reducible and irreducible representations of CCR has been recently proposed \cite{CW}. The goal of the present paper is to show the freedom of choosing different representations of CCR is at the roots of yet another controversy. The problem of `entanglement with vacuum' has led to the discussion whose never-ending character seems to be related to the fact the main question has never been precisely formulated. Its most recent manifestation is the paper \cite{Enk}, which will be the basis of our analysis. The physical configuration we discuss is the same as the one analyzed in \cite{Enk}. 

Consider two two-level atoms and a photon in superposition of two orthogonal modes. A Jaynes-Cummings type interaction couples each of these modes with a different atom. Initially the atoms are in a product state, but finally they become entangled. This conclusion can be reached without any assumption about the form of representation of the CCR algebra that models the modes:
\be
{[a_m,a_n^{\dag}]}=\delta_{mn}I_m,\label{CCR}
\ee
and $I_m=I_m^{\dag}$ commute with everything (i.e. belong to the center of the algebra). 
Now, what can be said about the entanglement present in the photon state? Apparently, and this is the conclusion reached in a number of papers (cf. \cite{Enk,G,AV}), the state had to be entangled because the interaction is a product of two local operations and thus cannot increase entanglement in the system. This is indeed true if one works with the irreducible representation constructed in terms of an infinite tensor product (we term it the 
`$\infty$-representation'). The $\infty$-representation is very popular in quantum optics community and is often treated as {\it the\/} representation of CCR. The problem is that this is only {\it a\/} representation and, according to the theorem of von Neumann \cite{Petz,BR}, there exists an infinite number of inequivalent irreducible representations of this algebra. Another representation can be found in the Berezin textbook \cite{Berezin}, so let us term it the  `$B$-representation'. Now the conclusion from \cite{G,AV,Enk} is no longer true: The state of the photon is not entangled, but the interaction is no longer local, and this is why the final state is entangled. We first show how the two calculations work in practice, and then perform a calculation in the reducible `finite $N$ representation' which, as shown for example in \cite{CW}, reconstructs the standard results with arbitrary precision for finite-time evolutions. Here the conclusion is again different: A single-mode state that is not entangled in the $\infty$-representation becomes entangled with vacuum in the $N$-representation if $N>1$. 

So the conclusion will be that the very notion of entanglement is well defined only after having specified the mathematical representation of a state and, in particular, the number of subsystems it describes. But the characterization such as a `two-mode state of a single photon' is not yet concrete enough to fix the mathematics of the problem. And this is the true source of the controversy.

\section{Calculation independent of representation}

Consider the Hamiltonian
\be
H &=&  R^{\dag}\otimes A+R\otimes A^{\dag}
=
\left(
\begin{array}{cc}
0 & A\\
A^{\dag} & 0
\end{array}
\right).
\ee
$A$ is an arbitrary operator, and $R$ is the atomic annihilation operator satisfying 
$R^2=0$, $R|+\rangle=|-\rangle$, $R^{\dag}|-\rangle=|+\rangle$, where $|+\rangle$ and $|-\rangle$ are the atomic excited and ground states, respectively. One can explicitly write the evolution operator without any need of specifying the form or algebraic properties of $A$:
\begin{widetext}
\be
e^{-iHt}
&=&
R^{\dag}R\otimes \cos \big(t \sqrt{AA^{\dag}}\big)
+
RR^{\dag}\otimes \cos \big(t \sqrt{A^{\dag}A}\big) 
-it
R^{\dag}\otimes \sinc \big(t \sqrt{AA^{\dag}}\big) A 
-it
R\otimes \sinc \big(t \sqrt{A^{\dag}A}\big)A^{\dag}.\label{1}
\ee
\end{widetext}
Here $\sinc x=(\sin x)/x$ and the square roots are well defined since both $AA^{\dag}$ and 
$A^{\dag}A$ are positive operators, no matter what $A$ one selects. Now take $R_1=R\otimes 1$, 
$R_2=1\otimes R$, and two annihilation operators $a_1$, $a_2$, satisfying the CCR algebra
$[a_m,a_n^{\dag}]=\delta_{mn}I_m$. The formula (\ref{1}) when applied to the Hamiltonian
$H=H_1+H_2$, 
\be
H_k &=& R_k^{\dag}\otimes i\,a_k-R_k\otimes i\,a_k^{\dag}, \quad k=1,2,
\ee
implies
$
e^{-iHt}
=
e^{-iH_1t}
e^{-iH_2t},
$
\begin{widetext}
\be
e^{-iH_kt}
&=&
R_k^{\dag}R_k\otimes \cos \big(t \sqrt{a_ka_k^{\dag}}\big)
+
R_kR_k^{\dag}\otimes \cos \big(t \sqrt{a_k^{\dag}a_k}\big) 
+t
R_k^{\dag}\otimes \sinc \big(t \sqrt{a_ka_k^{\dag}}\big) a_k 
-t
R_k\otimes \sinc \big(t \sqrt{a_k^{\dag}a_k}\big)a_k^{\dag}.
\ee
Let $|0\rangle$ be a normalized vector satisfying $a_k|0\rangle=0$, and let
$a=(a_1+a_2)/\sqrt{2}$. Then
\be
e^{-iHt}|-\rangle|-\rangle a^{\dag}|0\rangle
&=&
\frac{1}{\sqrt{2}}|-\rangle|-\rangle 
\Big(
\cos \big(t \sqrt{I_1}\big) a_1^{\dag}|0\rangle
+
\cos \big(t \sqrt{I_2}\big) a_2^{\dag}|0\rangle
\Big)\nonumber\\
&\pp=&+
\frac{t}{\sqrt{2}}
\Big(
|+\rangle|-\rangle \sinc \big(t \sqrt{I_1}\big) I_1|0\rangle
+
|-\rangle|+\rangle \sinc \big(t \sqrt{I_2}\big) I_2|0\rangle
\Big).\label{fs}
\ee
\end{widetext}
This final state is in general entangled, although the degree of entanglement is difficult to estimate at such an abstract level. But for irreducible representations the central elements $I_1$, $I_2$ are proportional to the identity $I$, and without loss of generality we can take $I_1=I_2=I$.
Then 
\be
e^{-iHt}|-\rangle|-\rangle a^{\dag}|0\rangle
&=&
\cos t |-\rangle|-\rangle a^{\dag}|0\rangle\\
&\pp=&+
\frac{\sin t}{\sqrt{2}}\Big(|+\rangle|-\rangle+|-\rangle|+\rangle \Big) |0\rangle.\nonumber
\ee
The reduced density matrix describing the two atoms will be important when we compare the predictions of irreducible representations with those of the reducible ones. It reads
\be
\rho
&=&
\cos^2 t
|-\rangle|-\rangle \langle-|\langle-| 
\label{rho}\\
&\pp=&+
\frac{1}{2}
\sin^2 t 
\Big(
|+\rangle|-\rangle 
+
|-\rangle|+\rangle
\Big)
\Big(
\langle+|\langle-|
+
\langle-|\langle+|
\Big).\nonumber
\ee
These formulas are valid for all irreducible representations of CCR. 
It is obvious that our state is maximally entangled at least at $t=\pi/2$, but even now, as we shall see in the second of the following sections, we cannot conclude that the initial state $a^{\dag}|0\rangle$ is entangled. 

\section{$\infty$-representation calculation}

The Hilbert space ${\cal H}_\infty$ is an infinite tensor product of single-oscillator Hilbert spaces: ${\cal H}_\infty={\cal H}^{\otimes\infty }$. The operators are 
\be
a_m &=& I^{\otimes(m-1)}\otimes a\otimes I^{\otimes \infty},
\ee
where $I$ is the identity in ${\cal H}$, and $[a,a^{\dag}]=I$. 
The vacuum is given by the infinite tensor product
\be
|\uu 0\rangle
&=&
|0\rangle\otimes\dots\otimes |0\rangle\otimes\dots =|0\rangle^{\otimes\infty}.
\ee
Such a representation has all the mathematical pathologies typical of infinite tensor products. If the index $n$ refers to a wave-vector in cavity, then a localized wavepacket must involve the entire infinite tensor product.  However, it is typical to reduce discussions of two-mode problems involving $a_1$ and $a_2$ to $a_1=a\otimes I$, $a_2=I\otimes a$, and the vacuum to $|\uu 0\rangle=|0\rangle|0\rangle$. Then $a_1^{\dag}|\uu 0\rangle=
|1\rangle|0\rangle$, $a_2^{\dag}|\uu 0\rangle=
|0\rangle|1\rangle$, and 
\be
a^{\dag}|\uu 0\rangle=
\frac{1}{\sqrt{2}}\Big(
|1\rangle|0\rangle
+
|0\rangle|1\rangle
\Big).
\ee
This state is explicitly entangled so there is basically nothing left to prove. 
The statement is made somewhat stronger by noting that the evolution operator in this representation is \cite{G,AV,Enk}
\be
e^{-iHt}
&=&
e^{t( R^{\dag}\otimes a- R\otimes a^{\dag})}
\otimes 
e^{t( R^{\dag}\otimes a- R\otimes a^{\dag})}.
\ee
We have ordered the Hilbert spaces in the tensor product so that the initial state can be written as
\be
\frac{1}{\sqrt{2}}|-\rangle|1\rangle|-\rangle|0\rangle 
+
\frac{1}{\sqrt{2}}|-\rangle|0\rangle|-\rangle|1\rangle.
\ee
$e^{-iHt}$
is a product of two operators that act {\it locally} on the Hilbert spaces of `the first atom plus the first oscillator', and `the second atom plus the second oscillator'. In such a system the entanglement can be exchanged only locally, and since the final state at $t=\pi/2$ involves nonlocal entanglement between the atoms, the initial entanglement of the oscillators had to be the same. This is consistent with the obvious constatation that initially the oscillators were entangled, and thus the state of light behind a beam splitter involved maximal entanglement between modes and vacuum.

\section{$B$-representation calculation}

Consider the Hilbert space of column vectors of the form \cite{Berezin}
\be
|\Psi\rangle
&=&
\left(
\begin{array}{c}
\Psi_0\\
\Psi_1(k_1)\\
\Psi_2(k_1,k_2)\\
\Psi_3(k_1,k_2,k_3)\\
\vdots
\end{array}
\right)
\ee
with the norm
\be
\langle\Psi|\Psi\rangle
&=&
\sum_{n=0}^\infty \int dk_1\dots dk_n  |\Psi_n(k_1,\dots,k_n)|^2.
\ee
We apply the convention where all the degrees of freedom of a photon are denoted by a single argument $k$, and $dk$ is an appropriate measure. 
We assume the functions are symmetric in their arguments. The Hilbert space is therefore the direct sum of symmetrized tensor products, ${\cal H}_B=\sum_{n=0}^\infty {\cal H}^{\otimes_Sn}$.
Denote by $P$ the projector on functions symmetric in the $k$-variables, and let the  functions $f_n$ form an orthonormal basis in the Hilbert space $\cal H$. The representation of CCR reads 
\begin{widetext}
\be
a_n^{\dag}
&=&
P\left(
\begin{array}{cccccc}
0 & \dots & & & &\\
0 & f_n(p_1) & 0 &\dots  & &\\
0 & 0      & \sqrt{2}\delta(p_1-k_1)f_n(p_2) & 0 & \dots&\\
0 & 0      & 0                             & \sqrt{3}\delta(p_1-k_1)\delta(p_2-k_2)f_n(p_3) & 0 &\dots\\
\vdots
\end{array}
\right)P.
\ee
\end{widetext}
$a_n$ and $a_n^{\dag}$ are related by Hermitian conjugation. It is understood that acting with $a_n^{\dag}$ on $|\Psi\rangle$ we perform integration over the $k$ variables. 

Let us now take two orthogonal functions, $f_1(p)$ and $f_2(p)$, representing the wavepackets leaving a beam splitter and localized at $t=0$ at its two sides. The vacuum state is
\be
|0\rangle
&=&
\left(
\begin{array}{c}
1\\
0\\
\vdots
\end{array}
\right).\label{Bvac}
\ee
Acting on the vacuum with $a^{\dag}=(a_1^{\dag}+a_2^{\dag})/\sqrt{2}$ we find
\be
a^{\dag}|0\rangle
&=&
\frac{1}{\sqrt{2}}
\left(
\begin{array}{c}
0\\
f_1(p_1)+f_2(p_1)\\
0\\
\vdots
\end{array}
\right)
\ee
and
\begin{widetext}
\be
e^{-iHt}|-\rangle|-\rangle a^{\dag}|0\rangle
&=&
\cos t |-\rangle|-\rangle 
\frac{1}{\sqrt{2}}
\left(
\begin{array}{c}
0\\
f_1(p_1)+f_2(p_1)\\
0\\
\vdots
\end{array}
\right)
+\sin t
\frac{1}{\sqrt{2}}\Big(|+\rangle|-\rangle+|-\rangle|+\rangle \Big) \left(
\begin{array}{c}
1\\
0\\
0\\
\vdots
\end{array}
\right).
\ee
\end{widetext}
The initial state was 
\be
|-\rangle|-\rangle 
\frac{1}{\sqrt{2}}
\left(
\begin{array}{c}
0\\
f_1(p_1)+f_2(p_1)\\
0\\
\vdots
\end{array}
\right)\in {\cal H}_A\otimes {\cal H}_A\otimes {\cal H}_B\label{B-ini}
\ee
and the final one, at $t=\pi/2$, is
\be
\frac{1}{\sqrt{2}}\Big(|+\rangle|-\rangle+|-\rangle|+\rangle \Big) \left(
\begin{array}{c}
1\\
0\\
0\\
\vdots
\end{array}
\right)\in {\cal H}_A\otimes {\cal H}_A\otimes {\cal H}_B,
\ee
where ${\cal H}_A$ is the Hilbert space of a two-level atom.
The dynamics is no longer a product one: The two atomic Hilbert spaces get coupled through 
${\cal H}_B$, and there is no natural splitting of the single-photon subspace ${\cal H}$ of
${\cal H}_B$ into tensor products. The initial state is explicitly a product state of  three subsystems. 

One can say that the same structure has the initial state in the $\infty$-representation, but the third subsystem is in an entangled state. This is true, but the problem with the $B$-representation is that the third subsystem does not have a natural decomposition into subsystems \cite{PH}. The distinction between the vacuum $|\uu 0\rangle$ and `a vacuum $|0\rangle$ at one side of the beam-splitter' is undefined in this representation \cite{?} since the vacuum is here unique, and corresponds to the vector (\ref{Bvac}). An entanglement with a unique vacuum must be trivial.
To the reader who is not yet convinced we propose an exercise: Try to prove that the initial state (\ref{B-ini}) is entangled by employing any known entanglement measure. It turns out that the exercise is simply ill posed.

\section{$N$-representation calculation}

The idea of the representation is to make the number of modes independent of the number of oscillators: Each oscillator is a wave-packet containing all the possible modes, and the number $N$ of oscillators is a parameter of the representation. In the weak limit $N\to\infty$ the $N$-representation reconstructs the results of irreducicle representations, but the formulas get automatically regularized in UV and IR regimes. Recently, in \cite{CW}, it was shown that the representation can be, in principle, directly tested in cavity QED. 
The $N$-representation of electromagnetic field operators was introduced in \cite{I}, and further analyzed and generalized in \cite{II,III,IV,V}. 

The representation is in the simplest case constructed as follows. Take an operator $a$ satisfying $[a,a^{\dag}]=1$ and the kets $| k\rangle$ corresponding to standing waves in some cavity. We define 
\be
a_k = | k\rangle\langle  k|\otimes a,\quad
I_k = | k\rangle\langle  k|\otimes 1.\label{I N=1}
\ee
The operators (\ref{I N=1}) satisfy $[a_k,a_{k'}^{\dag}]=\delta_{kk'}I_k$. $I_k$ is in the center of the algebra. The fact that $I_k$ is not proportional to the identity means that the representation is reducible. In our terminology this is the `$N=1$ representation'. Its Hilbert space $\cal H$ is spanned by the kets 
$| k,n\rangle=| k\rangle|n\rangle$, where $a^{\dag}a|n\rangle=n|n\rangle$. 
A vacuum of this representation is given by any state annihilated by all $a_k$. The vacuum state is not unique and belongs to the subspace spanned by $| k,0\rangle$. In our notation a $N=1$ vacuum state reads $|0\rangle=\sum_k O_k| k,0\rangle$ and is normalized by $\sum_k |O_k|^2=\sum_k Z_k=1$. The probabilities $Z_k=|O_k|^2$ play an essential role in the formalism, and of particular importance is their maximal value $Z=\max_k\{Z_k\}$. In relativistic theory  $Z$ is a Poincar\'e invariant that additionally parametrizes the representation and plays a role of renormalization constant. 
For $N\geq 1$ the representation space is given by the tensor power 
$\uu{\cal H}={\cal H}^{\otimes N}$, i.e. we take the Hilbert space of $N$ (bosonic) harmonic oscillators. For large $N$ the effective parameter that controls the representation is the product $NZ$. Let $A: {\cal H}\to {\cal H}$ be any operator for $N=1$. 
We denote
$A^{(n)}=I^{\otimes (n-1)}\otimes A \otimes I^{\otimes (N-n)}$, 
$A^{(n)}: \uu{\cal H}\to \uu{\cal H}$, for $1\leq n\leq N$. For arbitrary $N$ the representation is defined by
\be
\uu a_k
&=&
\frac{1}{\sqrt{N}}\sum_{n=1}^N a_k^{(n)},\quad
\uu I_k = \frac{1}{N}\sum_{n=1}^N I_k^{(n)}, \label{IN}\\
{[\uu a_k,\uu a_{k'}^{\dag}]} &=& 
\delta_{kk'}\uu I_k.
\ee
The $N$-oscillator vacuum is the $N$-fold tensor power of the $N=1$ case, a kind of Bose-Einstein condensate consisting of $N$ wavepackets:
\be
|\uu 0\rangle=|0\rangle\otimes\dots\otimes|0\rangle=|0\rangle^{\otimes N}.\label{vacN}
\ee
Let us denote $\uu a_1=\uu a_{k_1}$, $\uu a_2=\uu a_{k_2}$.
In order to apply the formula (\ref{fs}) we first spectrally decompose the operators 
$\uu I_1=\uu I_{k_1}$, $\uu I_2=\uu I_{k_2}$ \cite{V},
\be
\uu I_1 &=& \sum_{s=0}^N \frac{s}{N} E_1(s),\quad
\uu I_2 = \sum_{s=0}^N \frac{s}{N} E_2(s).
\ee
The spectral projectors $E_n(s)$ are in the center of CCR. 
Defining $a_n(s)=\uu a_n E_n(s)$, 
\be
{[a_n(s),a_m(s')^{\dag}]}=(s/N)\delta_{mn}\delta_{ss'}E_m(s),
\ee
we rewrite (\ref{fs}) as 
\begin{widetext}
\be
e^{-iHt/\sqrt{Z}}|-\rangle|-\rangle {\cal N}\uu a^{\dag}|\uu 0\rangle
&=&
\sum_{s=0}^N
|-\rangle|-\rangle \cos \Big( t \sqrt{\frac{s}{NZ}}\Big) {\cal N}
a(s)^{\dag}|\uu 0\rangle
\nonumber\\
&\pp=&+
\sum_{s=0}^N
\sin \Big( t \sqrt{\frac{s}{NZ}}\Big) \sqrt{\frac{s}{N}}
\frac{{\cal N}}{\sqrt{2}}
\Big(
|+\rangle|-\rangle E_1(s)|\uu 0\rangle
+
|-\rangle|+\rangle E_2(s)|\uu 0\rangle
\Big).\label{fsN}
\ee
\end{widetext}
As before $\uu a=(\uu a_1+\uu a_2)/\sqrt{2}$, and 
${\cal N}$ is chosen to make ${\cal N}\uu a^{\dag}|\uu 0\rangle$ normalized:
\be
|{\cal N}|^2\langle\uu 0|\uu a\,\uu a^{\dag}|\uu 0\rangle
&=&
|{\cal N}|^2(Z_1+Z_2)/2=1,\label{norm}\\
\langle\uu 0|\uu a_k\uu a_k^{\dag}|\uu 0\rangle
&=&
\langle\uu 0|\uu I_k|\uu 0\rangle
=\langle 0|I_k|0\rangle
=Z_k.
\ee
The formulas look much more complicated than those from the irreducible representations, but they turn out to imply the same physics if one appropriately defines the limit $N\to\infty$. 
The limit cannot be performed directly (i.e. in the strong sense) at the level of (\ref{fsN}). It has to be computed in a weak sense, for example at the level of averages or the atomic reduced density matrix, which reads
\begin{widetext}
\be
\rho
&=&
\frac{1}{Z_1+Z_2}
\sum_{s=0}^N 
|-\rangle|-\rangle \langle-|\langle-| 
\cos^2 \Big( t \sqrt{\frac{s}{NZ}}\Big)
\frac{s}{N}
\Bigg[
\left(
\begin{array}{c}
N\\
s
\end{array}
\right)
Z_1^s(1-Z_1)^{N-s}
+
\left(
\begin{array}{c}
N\\
s
\end{array}
\right)
Z_2^s(1-Z_2)^{N-s}
\Bigg]
\nonumber\\
&{}&+
\frac{1}{Z_1+Z_2}\sum_{s=0}^N
\sin^2 \Big( t \sqrt{\frac{s}{NZ}}\Big) \frac{s}{N}
|+\rangle|-\rangle \langle+|\langle-|
\left(
\begin{array}{c}
N\\
s
\end{array}
\right)
Z_1^s(1-Z_1)^{N-s}
\nonumber
\\
&{}&+
\frac{1}{Z_1+Z_2}\sum_{s,s'=0}^N
\sin \Big( t \sqrt{\frac{s}{NZ}}\Big) \sqrt{\frac{s}{N}}
\sin \Big( t \sqrt{\frac{s'}{NZ}}\Big) \sqrt{\frac{s'}{N}}
|+\rangle|-\rangle \langle-|\langle+| 
\left(
\begin{array}{c}
N\\
s,s'
\end{array}
\right)
Z_1^sZ_2^{s'}(1-Z_1-Z_2)^{N-s-s'}
\nonumber
\\
&{}&+
\frac{1}{Z_1+Z_2}\sum_{s,s'=0}^N
\sin \Big( t \sqrt{\frac{s}{NZ}}\Big) \sqrt{\frac{s}{N}}
\sin \Big( t \sqrt{\frac{s'}{NZ}}\Big) \sqrt{\frac{s'}{N}}
|-\rangle|+\rangle \langle+|\langle-|
\left(
\begin{array}{c}
N\\
s,s'
\end{array}
\right)
Z_1^sZ_2^{s'}(1-Z_1-Z_2)^{N-s-s'}
\nonumber
\\
&{}&+
\frac{1}{Z_1+Z_2}\sum_{s=0}^N
\sin^2 \Big( t \sqrt{\frac{s}{NZ}}\Big) \frac{s}{N}
|-\rangle|+\rangle \langle-|\langle+|
\left(
\begin{array}{c}
N\\
s
\end{array}
\right)
Z_2^s(1-Z_2)^{N-s}.\label{rhoN}
\ee
The law of large numbers for multinomial distributions implies
\be
\rho_\infty
&=&
\lim_{N\to\infty}
\rho\nonumber\\
&=&
\Bigg[
\frac{Z_1}{Z_1+Z_2}\cos^2 \Big( t \sqrt{\frac{Z_1}{Z}}\Big)
+
\frac{Z_2}{Z_1+Z_2}\cos^2 \Big( t \sqrt{\frac{Z_2}{Z}}\Big)
\Bigg]
|-\rangle|-\rangle \langle-|\langle-| 
\nonumber\\
&{}&+
\frac{Z_1}{Z_1+Z_2}
\sin^2 \Big( t \sqrt{\frac{Z_1}{Z}}\Big)
|+\rangle|-\rangle \langle+|\langle-|
+
\frac{Z_2}{Z_1+Z_2}\sin^2 \Big( t \sqrt{\frac{Z_2}{Z}}\Big) 
|-\rangle|+\rangle \langle-|\langle+|
\nonumber
\\
&{}&+
\sqrt{\frac{Z_1}{Z_1+Z_2}}\sin \Big( t \sqrt{\frac{Z_1}{Z}}\Big)
\sqrt{\frac{Z_2}{Z_1+Z_2}} \sin \Big( t \sqrt{\frac{Z_2}{Z}}\Big)
\Big(|+\rangle|-\rangle \langle-|\langle+| 
+
|-\rangle|+\rangle \langle+|\langle-|
\Big)\label{rhoNN}
\ee
\end{widetext}
Various arguments based on comparison of fields quantized in $N$-representations with those quantized by means of irreducible representations suggest that the vacuum probability distribution $Z_k$ is qualitatively of the form shown in Fig.~1. If the two wave vectors $k_1$, $k_2$ corresponding to $\uu a_1$ and $\uu a_2$ are assumed to belong to the plateau region 
(i.e.  $Z_1=Z_2=Z$) then $\rho_\infty$ becomes identical to (\ref{rho}). The dynamics looks as if it were regularized by the cut-off function $\chi_k=\sqrt{Z_k/Z}$ if $Z_k$ are in the IR or UV regimes. Let us also note that the dynamics is not generated by $H$ but by its renormalized version $H/\sqrt{Z}$, a property equivalent in quantum optics to bare charge renormalization.
\begin{figure}
\includegraphics[scale=0.55]{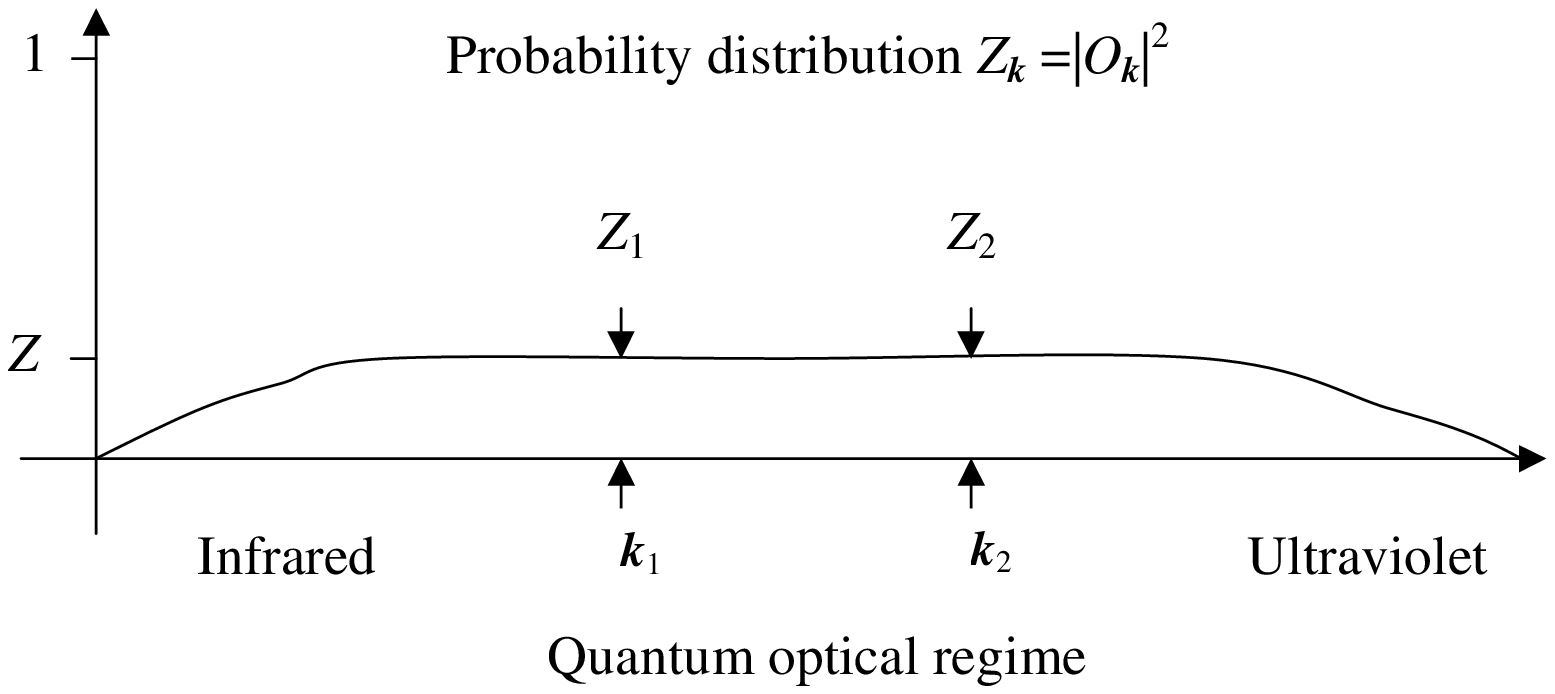}
\caption{Qualitative properties of the the vacuum probability distribution corresponding to the vacuum wave function in the $N$-representation. For wave vectors in the `quantum optics regime' we expect $Z_k/Z\approx 1$, where $Z$ is the maximal value of $Z_k$.}
\end{figure}
Internal consistency of calculations based on $N$-representations requires that $N<\infty$. Simultaneously, as discussed in detail in \cite{CW}, for any finite-time evolution one can choose such an $N$ that the evolution will be indistinguishable, within some given error bars, from the result based on irreducible representations. Taking $t=\pi/2$ one can always choose a finite $N$ guaranteeing that  (\ref{rhoN}) is arbitrarily close to the maximally entangled atomic state (\ref{rho}). Paradoxically, we have this agreement for a {\it large\/}  $N$, and not for $N=2$ which one might naively expect on the basis of the experiences with 
the $\infty$-representation. 

Concerning the entanglememnt with vacuum of the initial state, the $N$-representation reads explicitly
\be
\uu a^{\dag}|\uu 0\rangle
&=&
\frac{1}{\sqrt{N}}
\Big(
a^{\dag}|0\rangle|0\rangle\dots|0\rangle+\dots+|0\rangle\dots|0\rangle a^{\dag}|0\rangle
\Big).\nonumber
\ee
Here $a=(a_1+a_2)/\sqrt{2}$ with $a_1$ and $a_2$ defined for $N=1$. This state is highly entangled, but it would be entangled also for a single-mode problem, since
\be
\uu a_1^{\dag}|\uu 0\rangle
&=&
\frac{1}{\sqrt{N}}
\Big(
a_1^{\dag}|0\rangle|0\rangle\dots|0\rangle+\dots+|0\rangle\dots |0\rangle a_1^{\dag}|0\rangle
\Big)\nonumber
\ee
even though its $\infty$-representation is a product state. $\uu a_1^{\dag}|\uu 0\rangle$ is similar to the $\infty$-representation state
\be
\frac{1}{\sqrt{N}}
\Big(
|1\rangle|0\rangle\dots|0\rangle+\dots+|0\rangle\dots |0\rangle|1\rangle
\Big)\nonumber
\ee
representing a superposition of $N$ different modes, but is `more entangled' since each $N=1$ component $a_1^{\dag}|0\rangle$ is a superposition of all the wave-vector degrees of freedom. 

\section{Conclusions}

It is not clear which representation of CCR should be used to model quantum electromagnetic fields. The von Neumann theorem guarantees that there exists an infinite number of different irreducible representations, and we have not found a {\it single\/} quantum optics problem that required the use of the $\infty$-representation and not, say, the Berezin one. Even the most spectacular problems involving entanglement, such as teleportation or violation of Bell's inequalities, turned out to be solvable at a {\it representation independent\/} level. To the best of our knowledge all the formulas given in the recent reviews on linear optics quantum computation \cite{L1,L2} can be translated into a representation independent formalism, and then into the $B$-representation, say. Accordingly, it is absolutely irrelevant which irreducible representation we select for our explicit calculations. These remarks apply, in particular, to all the papers on entanglement with vacuum and nonlocality of a single photon where the calculations were performed in the 
$\infty$-representation \cite{1,2,3,4,5,6,7,8,9} (see however \cite{10}). No single experimental prediction would be affected by switching to the $B$-representation, but the discussion of entanglement would be changed, of course.

The paper \cite{GHZ} whose intention was to clarify the status of entanglement with vacuum has increased the confusion by making a distinction between `entanglement in configuration space' and `entanglement in Fock space', but the Fock space was identified with the $\infty$-representation. It is evident that any irreducible representation of CCR leads to the notion of a Fock space, i.e. a direct sum of Hilbert spaces constructed from a vacuum state by means of creation operators. The conclusions of \cite{GHZ} would not be possible to formulate if the authors discussed the problem in the $B$-representation of the Fock space. 

The situation changes if we consider the reducible representations. As opposed to irreducible representations here even a single-mode state of light is entangled with vacuum if $N>1$. The $N$-representation is in principle experimentally testable \cite{CW}. If we managed to determine the physical representation (representations?) of CCR, the discussion of entanglement in optical systems might finally become physically well posed.

\acknowledgments
We are indebted to M. Wilczewski and P. Horodecki for discussions and comments. An exchange of emails with L. Vaidman and S. J. van Enk allowed us to see the problem in a wider context, and we are grateful for their critical remarks. This work was done as a part of the Polish Ministry of Scientific Research and Information Technology (solicited) project PZB-MIN 008/P03/2003.


\begin{thebibliography}{99}
\bibitem{?}One cannot treat it as an argument against the $B$-representation. Formally, an identical situation occurs for any single quantum particle interacting with a system of detectors (a photographic plate, say; a two-level atom is the usual model of a photodetector), or a particle in a two-path interferometer where the two paths involve interactions with different devices (e.g. mirrors).
\end{thebibliography}
\end{document}